# Why there is no noon-midnight red shift in the GPS

Neil Ashby
Department of Physics, University of Colorado, Boulder
NIST Affiliate
email: ashby@boulder.nist.gov

Marc Weiss
Time & Frequency Division, National Institute of Standards and Technology
Boulder, CO
email: mweiss@boulder.nist.gov

**Abstract:** Although the effects of solar (and lunar) gravitational potentials on the frequencies of orbiting Global Positioning System (GPS) clocks are actually no more than a few parts in $10^{15}$, a naïve calculation appears to show that such effects are much larger, and depend on whether the orbiting clock is between the earth and the sun, or on the side of the earth opposite to the sun. Consequently questions about whether such effects have been properly accounted for in the GPS continue to arise. This issue has been discussed in a misleading way in terms of cancellations arising from a second-order Doppler shift in the literature for almost 50 years. The purpose of this article is to provide a correct argument, based on fundamental relativity principles, so that one may understand in a simple way why the effects of external solar system bodies on orbiting or earth-bound clocks in the GPS are so small. The relativity of simultaneity plays a crucial role in these arguments.

## 1. Introduction

The Principle of Equivalence states that over a sufficiently small region of space and time—such as in an accelerating laboratory—the effect of acceleration cannot be distinguished from a real gravitational field due to mass. A rocket's acceleration **a** induces a gravitational field $-\mathbf{a}$ within the rocket that has the same physical effects as a gravitational field of strength $\mathbf{g} = -\mathbf{a}$. A downward acceleration $\mathbf{a} = \mathbf{g}$ of a freely falling elevator in earth's gravity field induces a gravitational field $-\mathbf{a} = -\mathbf{g}$ that cancels the real gravitational field strength $\mathbf{g}$, resulting in weightlessness within the elevator. Since the earth and its satellites are in free fall in the gravitational field of external solar system bodies—principally the sun—the acceleration of this system towards the sun induces a gravitational field that cancels the real field strength due to the external bodies. The residual effect of external masses is found in the tidal potential. The purpose of this article is to provide a detailed proof of this effect within the freely-falling, earth-centered inertial (ECI) frame that underlies GPS satellites.

There are two relativistic effects to be considered: the gravitational frequency shift due to a potential difference, and the second-order Doppler effect, or time dilation, of moving clocks. Two clocks at different heights near earth's surface and whose gravitational potentials differ by $\Delta\Phi$ will suffer a fractional frequency difference given by

$$\frac{\Delta f}{f} = \frac{\Delta\Phi}{c^2} \simeq \frac{gh}{c^2} \qquad (1)$$

where $\Delta\Phi$ is the gravitational potential difference, $h$ is the altitude difference, and $c$ is the speed of light, $c = 299,792,458 \text{ m/s}$. (We discuss explicitly how such frequency measurements are made in Section 3.) When formulas such as (1) are used to compare the frequencies of two clocks in GPS satellites, one between the earth and the sun (the "noon" side), and one on the opposite side of the earth (the "midnight" side), a huge frequency difference is found due to the solar potential difference. To estimate the size of such an effect, we calculate the field strength at earth's center of mass due to the sun:

$$g_\odot = \frac{GM_\odot}{R^2} \simeq 0.00594 \text{ m/s}^2 \qquad (2)$$

where $R = 1.495 \times 10^{11}$ m is approximately the earth-sun distance, $M_\odot$ is the sun's mass, and $G$ is the Newtonian gravitational constant. The distance between two GPS satellites on opposite sides of the earth is $2a$, where the nominal GPS satellite orbit radius is $a = 2.6562 \times 10^7$ m. Then

$$\frac{\Delta f}{f} \simeq \frac{g_\odot(2a)}{c^2} = 3.5 \times 10^{-12} \qquad (3)$$

This shift is orders of magnitude larger than the typical fractional frequency stability of GPS clocks, which can be smaller than $10^{-14}$ at one day (Oaks 2005). Such a large effect would have to be accounted for if it was actually present, but it is not observed. The large size of the result in (3) has led to erroneous claims that effects due to lunar and solar potentials have not been accounted for properly in the GPS, as well as to incorrect explanations of the physical basis for certain cancellations that remove the effects of (3). All too often the authors of the present article are asked to review such claims. Here we shall show in detail why such claims are erroneous. We will estimate the size of residual lunar and solar potential effects on moving clocks in the ECI frame in the vicinity of the earth.

The outline of this article is as follows. In Section 2, we review published discussions of this "noon-midnight" redshift that mistakenly attribute its disappearance to a second-order Doppler effect. In Sections 3 and 4 we show that the true reason for the effect's absence is intimately related to the relativity of simultaneity. A more detailed derivation based on the results of tensor transformations is given in Section 5. Estimates of the residual lunar and solar tidal effects are given in Section 6.

## 2. An incorrect argument based on the second-order Doppler effect.

Because we will ultimately be interested in estimating the actual effect of lunar and solar potentials, we need more precise expressions for these potentials. We will concentrate on the effect due to the sun, and approximate the sun's potential by making a Taylor expansion similar to the Legendre polynomial expansion of the potential due to a point charge. Let $\mathbf{R}_0$ be the radius vector from the sun to earth's center of mass, and let $\mathbf{r}$ be the vector from earth's center of mass to the point of observation, such as the position of a GPS satellite. The center of the sun is taken to be the origin of coordinates, and the center of the earth is the origin of a local freely falling frame—the ECI frame. The sun's potential at the satellite location is

$$\Phi_\odot = -\frac{GM_\odot}{|\mathbf{R}_0 + \mathbf{r}|} = -\frac{GM_\odot}{R_0} + \frac{GM_\odot \mathbf{R}_0 \cdot \mathbf{r}}{R_0^3} - GM_\odot \frac{3(\mathbf{R}_0 \cdot \mathbf{r})^2 - R_0^2 r^2}{2R_0^5} + ... \qquad (4)$$

The first term on the right side in (4) is the sun's potential at earth's center. It represents an average solar potential, felt by all clocks in the vicinity of earth, and affects the frequencies of all clocks the same way, so could not be detected by comparing the frequencies of two clocks near earth as estimated in (1-3), above. The coefficient of the vector $\mathbf{r}$ in the second term on the right in (4) describes the sun's gravitational field strength at earth's center. This term is responsible for the apparent frequency shift discussed in (3). The last term in (4) is the solar tidal potential, which falls off as the cube of the distance from the sun, grows quadratically with distance from earth's center, and depends on the angle between $\mathbf{R}_0$ and $\mathbf{r}$. More generally, if $\Phi$ is the total gravitational potential due to solar system bodies *other than the earth*, we can expand about the center of the earth at $\mathbf{R}_0$ and write

$$\Phi = \Phi_0 + \sum_{i=1}^{3}(X^i - R_0^i)\frac{\partial \Phi}{\partial X^i} + \frac{1}{2}\sum_{i,j=1}^{3}(X^i - R_0^i)(X^j - R_0^j)\frac{\partial^2 \Phi}{\partial X^i \partial X^j} + ... \qquad (5)$$

where $X^i = \{X, Y, Z\}$ are the Cartesian components of $\mathbf{R}$ and the derivatives are evaluated at $\mathbf{R}_0$. To simplify some results we shall write

$$\Phi_{tidal} = \frac{1}{2}\sum_{i,j=1}^{3}(X^i - R_0^i)(X^j - R_0^j)\frac{\partial^2 \Phi}{\partial X^i \partial X^j} \qquad (6)$$

It is the term linear in $X^i - R_0^i$ in (5) that leads to erroneous estimates of the "noon- midnight red shift." We show below that this effect is cancelled by a contribution arising from the relativity of simultaneity.

Some authors (Chubb 1994; Hoffman 1961) mistakenly attribute the cancellation to a second-order Doppler effect. To understand what is wrong with such an argument, we introduce a coordinate system whose origin is falling along with earth's center, but which has one axis--say the y-axis--always pointing towards the sun. (This is diagrammed in Figure 1.) We compare a clock at distance $R$ from the sun, with a clock at distance $R - y$ from the sun. Thus consider a clock on the $y$-axis at some fixed distance $y$ from the origin of these coordinates (which is at earth's center of mass), on the side toward the sun. The earth-sun distance is $R$, and we let the angular velocity of the earth about the sun be $\Omega$. Because the distance of the clock from the sun is $R - y$, its velocity is

$$V = \Omega(R - y) \qquad (7)$$

Relative to clocks at rest in the sun-centered coordinate system, the clock will suffer a frequency shift from the second-order Doppler effect (time dilation), in the amount

$$\frac{\Delta f}{f} = -\frac{1}{2}\frac{(\Omega(R-y))^2}{c^2} \simeq -\frac{1}{2}\frac{\Omega^2 R^2 - 2\Omega^2 Ry}{c^2} = -\frac{1}{2}\frac{\Omega^2 R^2}{c^2} + \frac{\Omega^2 Ry}{c^2} \qquad (8)$$

when $y \ll R$. Relative to a clock at the origin of the local frame, the frequency shift will be

$$\frac{\Delta f}{f} = -\frac{1}{2}\frac{\Omega^2 R^2}{c^2} + \frac{\Omega^2 Ry}{c^2} - \left(-\frac{1}{2}\frac{\Omega^2 R^2}{c^2}\right) = \frac{\Omega^2 Ry}{c^2} \qquad (9)$$

Note that the centripetal acceleration of the local frame's origin, at earth's center, is $\Omega^2 R$ towards the sun. The induced gravitational field due to this acceleration is $\Omega^2 R$ outwards.

The sun's gravitational effect on clock frequency needs to be calculated also. Relative to a clock at rest at distance $R$ from the sun, at the origin of the local frame, the gravitational fractional frequency shift of a clock at $y$ will be

$$\frac{\Delta f}{f} = \frac{1}{c^2}\left(\Phi(R-y) - \Phi(R)\right) = -\frac{GM_\odot}{c^2}\left(\frac{1}{R-y} - \frac{1}{R}\right) \simeq -\frac{GM_\odot y}{c^2 R^2} \qquad (10)$$

The sum of these two effects is

$$\frac{\Delta f}{f} = +\frac{\Omega^2 R y}{c^2} - \frac{GM_\odot y}{c^2 R^2} \qquad (11)$$

But if the origin of local coordinates is in free fall, the centripetal acceleration is supplied by the force of gravity, so

$$\Omega^2 R = \frac{GM_\odot}{R^2} \qquad (12)$$

and the terms linear in $y$ in (11) cancel. This would be true whether $y$ is positive or negative.

So it appears that there is a nice explanation for the cancellation that must be present to satisfy the Equivalence Principle. But the argument cannot be sustained, *because the axes of the locally inertial, earth-centered reference frame cannot always remain pointing toward the sun.* A freely falling, locally inertial reference frame *is not rotating*. Clocks at different fixed locations within that frame describe circles of equal radii in the same amount of time and can suffer no relative Doppler shift. The coordinate axes of the J2000 reference frame in which GPS clocks are synchronized point in fixed directions in the cosmos, except for tiny rotations (precession, nutation) that are negligible for this discussion. This means that as the earth and its satellites orbit around the sun, all points near the origin in the local frame move with equal velocities, and must all experience the same second-order time dilation relative to clocks at rest in the sun-centered coordinate system. So the velocity effect calculated in (3) does not exist. What effect is it then, which cancels the gravitational part of the frequency shift calculated in (11)?

Hoffman (1961) gives essentially the incorrect argument that we have just critiqued. We shall return to discuss this paper in more detail in a later section.

In their paper describing a satellite mission to measure the gravitational redshift, Kleppner et al. (1970) quote Hoffman while paraphrasing the implications of the principle of equivalence in a manner that parallels the argument given in the first paragraph of the introduction, above. Such arguments are correct but lack mathematical detail:

> The earth and the satellite are both in free fall towards the Sun. If we think of them as members of an isolated system, it is readily seen that their common acceleration in the Sun's field can have no observable effect. Thus the linear variation of the Sun's potential makes no contribution to the red shift, though higher order terms do contribute. This is a good example of the local nature of the equivalence principle: to the extent that curvature of the Sun's potential can be neglected, the Earth-Satellite system can be regarded as local.

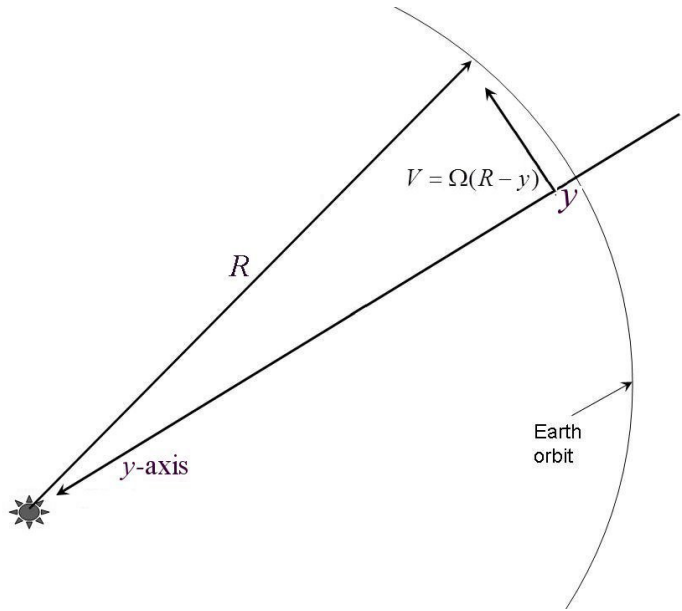

Figure 1. Local coordinates centered on the earth
with the $y-$axis pointing toward the sun.

     S. Chubb (1994) attempted to explain some small diurnal effects on GPS clocks in just the terms described in Sect. 2, above. Chubb claimed on the basis of the (incorrect) Doppler-field strength cancellation arguments, that Kleppner, Vessot and Ramsey's discussion of tidal potentials was incorrect, and a result of about the right order of magnitude was derived for the diurnal effects. The effects themselves later turned out to be due to poor station coordinates. Unfortunately, numerous algebraic mistakes were made in this paper (Chubb 1994), which invalidate those results.

     In the next section we shall show that the resynchronization of clocks in the freely falling system--that is, the relativity of simultaneity, rather than a second-order Doppler effect--is actually responsible for enforcing the equivalence principle.

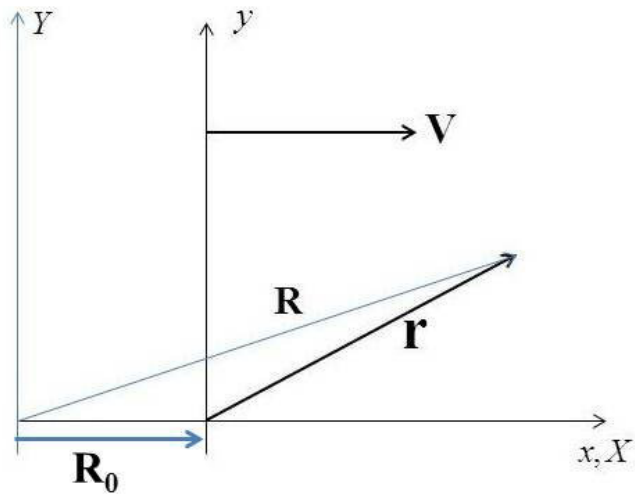

Figure 2.  Ideal model of synchronization process.

## 3. Equivalence Principle and the relativity of simultaneity

In this section we demonstrate the connection between the Equivalence Principle and the relativity of simultaneity.  Our method is to generalize the Lorentz transformations to the case of an accelerated frame of reference, wherein a gravitational field arises, then apply the result to the proper time elapsed on an accelerating clock.  There are several relativistic effects caused by earth's rotation, which can obscure the point we are trying to make, so we shall momentarily treat the earth as a non-rotating point mass.  Earth rotation will be discussed later.

We first show that the relativity of simultaneity is a consequence of the constancy of the speed of light.  Figure 2 illustrates the process of synchronization of an earth-orbiting clock from a monitor station.  (This is an ideal model that ignores earth rotation.) The displacement of the clock from the monitor station is $\mathbf{r}$ in the local, earth-centered, freely falling inertial frame.  We use upper case letters $X, Y$ to denote reference axes centered on the sun.  The vector from the sun to the satellite clock is denoted by $\mathbf{R}$.  The local (moving) frame has velocity $\mathbf{V}_0$ relative to the sun-centered frame.  A synchronization signal sent from the origin of local coordinates at $\mathbf{R}_0$ propagates with the speed $c$ to the satellite vehicle (SV) clock, taking an amount of time $\Delta t$ such that

$$|\mathbf{r}| = c\Delta t \qquad (13)$$

We neglect terms of second order in the small quantity $V/c$ --that is, we neglect Lorentz contraction and time dilation, and seek to calculate the propagation time in the sun-centered system. The principle of the constancy of the speed of light implies that this time is

$$\Delta T = \frac{1}{c}|\mathbf{R} - \mathbf{R}_0| \tag{14}$$

where $\mathbf{R}_0$ is the position of the transmission event in the sun-centered frame. Gravitational time delays are only a few hundred picoseconds and we neglect them here. (Pétit & Wolf 1994) The simplest possible transformation, of first order in the velocity, between the sun-centered system and the earth-centered system is a Galilean transformation for space:

$$\mathbf{R} = \mathbf{r} + \mathbf{V}t + const \tag{15}$$

where the constant depends on the synchronization convention at some initial instant. For example, if the clock at the origin of local coordinates is set to $t_0$ when the synchronization signal is transmitted, we would write

$$\mathbf{R} = \mathbf{R}_0 + \mathbf{r} + \mathbf{V}(t - t_0) = \mathbf{R}_0 + \mathbf{r} + \mathbf{V}\Delta t.$$

Thus to leading order, for the synchronization signal

$$\Delta T = \frac{1}{c}|\mathbf{R} - \mathbf{R}_0| = \frac{1}{c}|\mathbf{r} + \mathbf{V}\Delta t| \simeq \frac{1}{c}|\mathbf{r}| + \frac{\mathbf{V} \cdot \mathbf{r}\Delta t}{c|\mathbf{r}|}$$
$$= \Delta t + \frac{\mathbf{V} \cdot \mathbf{r}}{c^2} \tag{16}$$

where we have used (13). Thus we obtain

$$\Delta T = \Delta t + \frac{\mathbf{V} \cdot \mathbf{r}}{c^2}$$

The simplest possible time transformation, correct to first order in the velocity, between the two coordinate systems is therefore

$$T = t + \frac{\mathbf{V} \cdot \mathbf{r}}{c^2} + const \tag{17}$$

This expresses the relativity of simultaneity; it is thus a fundamental consequence of constancy of speed of light.

We can now demonstrate the connection between the Equivalence Principle and the relativity of simultaneity. We generalize the Lorentz transformations to the case of an accelerated frame of reference, wherein a gravitational field arises, then apply the result to the proper time elapsed on a freely falling, accelerated clock.

In preparation for development of a coordinate transformation from one centered on the sun to a freely falling coordinate system in orbit around the sun, we quote to leading order the Lorentz transformations from one inertial frame to another. Let $(cT, X, Y, Z)$ represent space-time coordinates in one inertial frame and $(ct, x, y, z)$ represent coordinates in a second inertial frame that moves uniformly with respect to the first with relative speed $V$ along the mutually parallel $x, X$-axes. Keeping only terms linear in velocity, these transformations are

$$X = x + Vt = x + \frac{V}{c}(ct);$$
$$Y = y; \quad Z = z; \quad (18)$$
$$cT = ct + \frac{V}{c}x$$

Here we are multiplying the time with the speed of light to give a time variable having units of length. Constants added to the right sides of (17) reflect initial conditions. They do not affect the following arguments. The last term in (17) expresses relativity of simultaneity. It means that clocks at rest at different positions $x$ in one frame, which have been synchronized using constancy of the speed of light, will not be synchronized in another inertial frame moving parallel to the $x-$ axis. This effect is of first order in velocity.

As the earth orbits around the sun, earth's velocity continually changes. We can describe relativistic effects by introducing a succession of instantaneous inertial frames, each with its origin at earth's center, and having velocity **V** equal to earth's instantaneous velocity, and maintaining axes in fixed directions relative to the stars. In each one of these instantaneous inertial frames, we can imagine that the process of synchronization of a GPS clock is continually being repeated, or is being repeated as often as necessary. The last term in (18) becomes

$$\frac{\mathbf{V} \cdot \mathbf{r}}{c}$$

where **V** is a slowly changing function of the time $T$. Only the component of the observation point **r** that is parallel to the velocity enters the time transformation. The origin of each of these instantaneous inertial frames is accelerated.

In general relativity the increment of proper time $d\tau$ on a clock is expressed through the metric tensor, which is a solution of the Einstein field equations. The scalar invariant proper time is given approximately by (Martin 1996; Schutz 1990):

$$c^2 d\tau^2 = \left(1 + \frac{2\Phi_t}{c^2}\right) c^2 dT^2 - \left(1 - \frac{2\Phi_t}{c^2}\right)\left(dX^2 + dY^2 + dZ^2\right) \quad (19)$$

where $\Phi_t$ is the total gravitational potential due to all solar system bodies.

Since we shall be discussing the fractional frequency offset of one clock from another, we here describe in some detail the meaning of such an offset. Atomic reference clocks such as those based on the hyperfine structure of $Cs^{133}$ provide time and frequency measurements by comparing the number of cycles $N$ elapsed on the clock with the number of cycles $N_{ref}$ of the reference oscillator, and this $N_{ref}$ determines the measurement interval. Cycles are relativistic scalar invariants. If $f_{ref}$ is the defined frequency of the reference then the time elapsed on the reference is $t_{ref} = N_{ref} / f_{ref}$. The measured frequency of a co-located clock will be $f = N / t_{ref} = Nf_{ref} / N_{ref}$. The fractional frequency shift of the co-located clock is then $f / f_{ref} - 1 = N / N_{ref} - 1$.

The frequency, $f_{ref}$, of the coordinate time variable that appears in the metric, and in the field equations of General Relativity, is realized in the GPS by applying relativistic corrections to orbiting atomic clocks. These clocks corrected for rate can be self-consistently synchronized to produce a time scale in the ECI frame; they beat at a rate defined by earth-fixed reference clocks on earth's geoid. GPS coordinate time is thus realized as a reference out to and beyond the orbits of geosynchronous satellites. A measurement of time interval using GPS coordinate time then has meaning over the entire space and can be measured using coordinate clocks at different locations. The proper time interval on a clock moving from one point to another can then be measured using GPS coordinate time as

$$\tau = t_{ref} N / N_{ref}.$$

The fractional frequency shift offset from the coordinate rate can then be expressed as

$$f / f_{ref} - 1 = \tau / t_{ref} - 1.$$

Note that we are not considering here the effect of first-order Doppler shifts such as those occurring when signals are exchanged between relatively moving clocks.

The variable $T$ in (19) is the barycentric coordinate time, a single time variable that is meaningful over the entire region in which (19) is a valid (approximate solution) of Einstein's field equations. The coordinate time difference $dT$ between two events corresponding to a proper time difference $d\tau$ on a clock is obtained by substituting directly into (19); obviously motion of the clock through space will come in to this relationship. Clocks near earth move slowly relative to the sun-centered frame; their velocities are small compared to that of light, so for the problem at hand we can set

$$dX^2 + dY^2 + dZ^2 = \mathbf{V}^2 dT^2 = \frac{V^2}{c^2}(c^2 dT^2) \qquad (20)$$

where $\mathbf{V}$ is the velocity of the clock. The product $2\Phi V^2/c^4$ is extremely tiny (of order $10^{-18}$) and can be neglected. Then keeping only leading terms of order $c^{-2}$, the proper time becomes

$$c^2 d\tau^2 = \left(1 + \frac{2\Phi}{c^2} - \frac{V^2}{c^2}\right) c^2 dT^2 \tag{21}$$

Taking square roots and solving for $cdT$ while keeping only leading terms,

$$cdT = \left(1 - \frac{\Phi}{c^2} + \frac{1}{2}\frac{V^2}{c^2}\right) cd\tau \tag{22}$$

We now construct a transformation from solar system coordinates to locally inertial, freely falling coordinates by accounting for several important relativistic effects. The time scale of the local inertial coordinates differs from the time $T$ because a clock on earth suffers from time dilation and gravitational frequency shifts that are variable due to the eccentricity of earth's orbit; this is already represented by the correction terms in (22). The earth's potential and rotation also impact the transformation; this will be discussed later.

We imagine a clock at earth's center whose proper time $\tau_0$ is the reference for the local time scale. Then integrating (22),

$$cT = \int_0^\tau \left(1 - \frac{\Phi_0}{c^2} + \frac{1}{2}\frac{V_0^2}{c^2}\right) cd\tau_0 \tag{23}$$

where $\mathbf{V}_0 = \dot{\mathbf{R}}_0$ is the velocity of the reference position and where $\Phi_0$ does not include earth's gravitational potential—it is the potential at earth's center due to other solar system bodies. A technical point is that there is no such clock, and we haven't yet accounted for the earth's potential and rotational motion in this expression.

We also need to account for other relativistic effects of order $c^{-2}$. Lorentz contraction parallel to the direction of motion, together with a possible length scale change due to external potentials, can be expressed by the transformation

$$X^k = R_0^k(T) + x^k \left(1 + \frac{S}{c^2}\right) + \frac{1}{2}\frac{V_0^k (\mathbf{V}_0 \cdot \mathbf{r})}{c^2} \tag{24}$$

where $R_0^k(T)$ is the position of earth's center, a function of the time. The last term in (24) is just Lorentz contraction parallel to the relative velocity (Ashby & Bertotti 1986). The Latin index $k$ runs from 1 to 3 and labels spatial components. The scale change correc-tion, $S/c^2$, in (24) is included for completeness but it will be seen that only the leading terms in (24) are needed.

Combining the relativity of simultaneity, (17), with the reference time scale, (23), and the transformation of spatial coordinates, (24), we propose to consider these transformation equations:

$$T = \int_0^\tau \left(1 - \frac{\Phi_0}{c^2} + \frac{1}{2}\frac{V_0^2}{c^2}\right) d\tau_0 + \frac{\mathbf{V}_0 \cdot \mathbf{r}}{c^2} \tag{25}$$

$$X^k = R_0^k(T) + x^k\left(1 + \frac{S}{c^2}\right) + \frac{1}{2}\frac{V_0^k(\mathbf{V}_0 \cdot \mathbf{r})}{c^2} \tag{26}$$

The new coordinates in the freely falling frame will be $\{\tau = t, x, y, z\}$. In general relativity, any non-singular coordinate transformation is permissible; it will be shown that the transformations of (25-26) greatly simplify the metric in the ECI frame.

## 5. Transformation to the local freely-falling frame.

The proper time of an earth-orbiting clock, (19), is to a sufficient approximation

$$d\tau^2 = \left(1 + \frac{2(\Phi + \Phi_e)}{c^2}\right) dT^2 - \frac{1}{c^2}\left(dX^2 + dY^2 + dZ^2\right) \tag{27}$$

Where the total gravitational potential has been separated into a contribution from the earth, $\Phi_e$, and a contribution from other bodies, $\Phi$. We carry along earth's potential to show that it does not affect the cancellations arising from relativity of simultaneity. The procedure is to express this scalar proper time in terms of local coordinates, (18), using the transformations of (25-26). We first need the increment $dT$ from (25), keeping in mind that the relative velocity $\mathbf{V}_0$ is a function of time. Thus differentiating $T$,

$$dT = \left(1 - \frac{\Phi_0}{c^2} + \frac{1}{2}\frac{V_0^2}{c^2}\right) d\tau_0 + \frac{1}{c^2}\left(d\mathbf{V}_0 \cdot \mathbf{r} + \mathbf{V}_0 \cdot d\mathbf{r}\right) \tag{28}$$

In the last term in (28) we can multiply and divide the last two terms by $d\tau_0$ since these terms already have a very small factor ; i.e.,

$$d\mathbf{V}_0 = (d\mathbf{V}_0/dT)dT \simeq (d\mathbf{V}_0/d\tau_0)d\tau_0 \simeq \mathbf{A}_0 d\tau_0.$$

We obtain

$$dT = \left(1 - \frac{\Phi_0}{c^2} + \frac{1}{2}\frac{V_0^2}{c^2} + \frac{\mathbf{A}_0 \cdot \mathbf{r}}{c^2} + \frac{\mathbf{V}_0 \cdot \mathbf{v}}{c^2}\right) d\tau_0 \tag{29}$$

where $\mathbf{v}$ is the clock velocity relative to the origin of the local frame. Here the acceleration of the origin of the local ECI frame is $\mathbf{A}_0 = d\mathbf{V}_0 / dT \simeq d\mathbf{V}_0 / dt$. Also,

$$\begin{aligned}dX^k &= \frac{dR_0^k}{dT} dT + dx^k \left(1 + \frac{S}{c^2}\right) + \frac{1}{2}\frac{V_0^k \mathbf{V}_0 \cdot d\mathbf{r}}{c^2} \\ &+ x^k \frac{dS}{c^2 dT} dT + \frac{1}{2}\frac{dV_0^k \mathbf{V}_0 \cdot d\mathbf{r} + V_0^k d\mathbf{V}_0 \cdot d\mathbf{r}}{c^2}.\end{aligned} \tag{30}$$

To compute the spatial part of the interval in (27), there is already a factor of $c^{-2}$ in front so keeping only terms of this order, all the relativistic corrections in (30) can be neglected. The spatial contributions are

$$\begin{aligned}-\frac{1}{c^2}(dX^2 + dY^2 + dZ^2) &= -\frac{1}{c^2} \sum_{i,j=1}^{3} \delta_{ij} \left(V_0^i dT + dx^i\right)\left(V_0^j dT + dx^j\right) \\ &\simeq -\frac{1}{c^2} \sum_{i,j=1}^{3} \delta_{ij} \left(V_0^i d\tau_0 + \frac{dx^i}{d\tau_0} d\tau_0\right)\left(V_0^j d\tau_0 + \frac{dx^j}{d\tau_0} d\tau_0\right) \\ &= -\frac{(\mathbf{V}_0 + \mathbf{v})^2}{c^2} d\tau_0^2\end{aligned} \tag{31}$$

The square of $dT$ by itself is, to the order of the calculation from (29),

$$dT^2 = \left(1 - \frac{2\Phi_0}{c^2} + \frac{V_0^2}{c^2} + \frac{2\mathbf{A}_0 \cdot \mathbf{r}}{c^2} + \frac{2\mathbf{V}_0 \cdot \mathbf{v}}{c^2}\right) d\tau_0^2 \tag{32}$$

Now we insert the expansion, (5), and (31) and (32) into (27) and expand.

$$\begin{aligned}d\tau^2 &= \left(1 + \frac{2}{c^2}\left(\Phi_0 + \sum_{i=1}^{3}(X^i - X_0^i)\frac{\partial \Phi}{\partial X^i} + \Phi_{tidal}\right) + \frac{2\Phi_e}{c^2}\right) \\ &\times \left(1 - \frac{2\Phi_0}{c^2} + \frac{V_0^2}{c^2} + \frac{2\mathbf{A}_0 \cdot \mathbf{r}}{c^2} + \frac{2\mathbf{V}_0 \cdot \mathbf{v}}{c^2}\right) d\tau_0^2 - \frac{(\mathbf{V}_0 + \mathbf{v})^2}{c^2} d\tau_0^2\end{aligned} \tag{33}$$

Obviously the terms in $\Phi_0$ cancel. To the order of the calculation, the distance $(X^i - X_0^i) = r^i$. Also, the terms in $V_0^2$ cancel as do cross terms of the form $\mathbf{V}_0 \cdot \mathbf{v}$. We expand and regroup the remaining terms:

$$d\tau^2 = \left(1 + \frac{2\Phi_e + 2\Phi_{tidal}}{c^2} + \frac{2\mathbf{r}\cdot\nabla\Phi}{c^2} + \frac{2\mathbf{A}\cdot\mathbf{r}}{c^2} - \frac{v^2}{c^2}\right)d\tau_0^2. \tag{34}$$

If the origin of local coordinates is in free fall, the acceleration is caused by the gravitational force, which is the negative of the gradient of the potential:

$$\mathbf{A} = -\nabla\Phi. \tag{35}$$

The linear terms in $\mathbf{r}$ cancel. Tracing back where the acceleration term comes from, it arises from the time derivative of the velocity term in the transformation, (25). This shows how the equivalence principle applies, through the relativity of simultaneity.

The remaining terms represent the contributions due to the gravitational potential of the earth itself, and time dilation due to motion of the clock relative to the earth's center. One further transformation is required to account for the fact that the earth's geoid is a surface of constant gravitational potential in the earth-fixed rotating frame and all clocks at rest on this rotating surface beat at the same rate. We therefore introduce one further time transformation:

$$\tau_0 = (1 - L_G)t, \tag{36}$$

where

$$L_G = 6.969290134 \times 10^{-10}. \tag{37}$$

The variable $t$ corresponds to TAI. $L_G$ is a defined constant (McCarthy 2003) that accounts for earth's multipole gravitational potential, as well as time dilation due to rotation, on earth's rotating geoid. Because $L_G$ is constant, it has no impact on the calculation of acceleration terms. The above derivation is valid whether or not the orbits of earth or its satellites are circular.

In an Appendix, Hoffman (1961) gives a sequence of four coordinate transformations that approximately reduce the metric, considering the earth as a point mass, to a Minkowski metric in the neighborhood of an object falling in a circular orbit of radius $R$ around the mass. Hoffman's net time transformation, keeping only terms of order $c^{-2}$, and in our notation, is

$$T = \tau\left(1 + \frac{GM}{c^2 R} + \frac{\omega^2 R^2}{2c^2} - \frac{\omega^2 Rx}{c^2}\right) + \frac{\omega Ry}{c} \tag{38}$$

For comparison we write out (25) for a circular orbit in the $x-y$ plane:

$$T = \tau\left(1 + \frac{GM}{c^2 R} + \frac{\omega^2 R^2}{2c^2}\right) + \frac{\omega R y}{c^2}\cos(\omega T) - \frac{\omega R x}{c^2}\sin(\omega T) \quad (39)$$

The time dependence in (39) is evident. For $\omega T \ll 1$, the last term in (39) resembles the last term in parentheses in (38), which is the term responsible for cancellation of the noon-midnight redshift contributions and which is described by Hoffman as a second-order Doppler effect. The coordinate transformations discussed by Hoffman do not give the correct tidal potential contributions.

## 6. Estimates of the lunar and solar tidal redshift contributions.

The frequency shift of an orbiting clock due the moon's tidal potential—the last term in (4), for example, can quickly be estimated for a satellite in a circular orbit of radius $r \sim 2.65 \times 10^7$ meters in the ecliptic plane. Relative to a clock at earth's center,

$$\frac{\Delta f}{f} = -\frac{GM_M r^2}{c^2 R^3}\left(\frac{3(\cos\varphi)^2 - 1}{2}\right) \quad (40)$$

where $M$ is the moon's mass, $\varphi$ is the angle between the vector from the moon to earth's center, and the vector from earth's center to the satellite clock. The coefficient in front of the parentheses in (40) is

$$-\frac{GM_M r^2}{c^2 R^3} \sim 6.7 \times 10^{-16} \quad (41)$$

In addition, tidal perturbations themselves cause time-varying changes in the orbit radius, which causes changing contributions of the same order of magnitude in the monopole contribution of the lunar potential to the frequency shift. Tidal perturbations also cause time-varying changes in the orbital velocity, which in turn causes changing contributions to the time-dilation contribution (second-order Doppler) to the frequency shift. The geometry of GPS satellite orbits is complex since the orbits are inclined with respect to earth's equator, which is itself inclined to the ecliptic; the result depends in a complicated way on specific orbital elements. The sun's tidal potential causes similar types of contributions that are about half as large. The main conclusion is that there are several significant contributions having a period of about 6 hours and a net amplitude of about $3.2 \times 10^{-15}$. The amplitudes of such tidal effects increase as the square of the orbital radii, so for clocks in the GALILEO system tidal effects on fractional frequency shifts of orbiting clocks will have maximum amplitudes of around $4 \times 10^{-15}$. For clocks in geosynchronous orbits, the amplitudes should be about $8 \times 10^{-15}$. These effects are currently being studied.

## 8. Summary

In summary, we have shown why the effect of the solar potential can be neglected in the GPS. It is a consequence of the principle of equivalence. The induced kinematical acceleration that arises when the reference frame is in free fall in a gravitational field, superimposes on the true gravitational field due to the sun's mass, in such a way as to precisely cancel. When working out the details of this phenomenon, the relativity of simultaneity plays a crucial role, since the required acceleration term arises from the time derivative of the term arising, in ordinary Lorentz transformations, from the relativity of simultaneity. This has been worked out in greater detail in the literature (Ashby & Bertotti 1986; Ashby & Saless 1988; Nelson 1994). We have applied these results to obtain estimates of the fractional frequency shifts due to tidal effects for clocks in earth-bound satellites.